\documentclass[journal=jctcce,manuscript=article]{achemso}

\usepackage{amsmath}
\usepackage{graphicx}
\usepackage{booktabs}
\usepackage{xcolor,colortbl}
\definecolor{literaturegray}{gray}{0.92}

\newcommand{\tm}{T_{\mathrm{m}}}
\newcommand{\tdm}{T_{\mathrm{dm}}}

\title{Training Dataset Matters for Machine Learning Interatomic Potentials: Water Simulation as an Example}

\author{Yifan Li}
\email{yifanl0716@gmail.com}
\altaffiliation{These authors contributed equally to this work.}
\affiliation{Department of Chemistry, Princeton University, Princeton, NJ 08544, USA}
\author{Bingjia Yang}
\altaffiliation{These authors contributed equally to this work.}
\affiliation{Department of Chemistry, Princeton University, Princeton, NJ 08544, USA}
\author{Chunyi Zhang}
\affiliation{Eastern Institute of Technology, Ningbo, Zhejiang 315200, China}
\author{Axel Gomez}
\affiliation{Department of Chemistry, Princeton University, Princeton, NJ 08544, USA}
\author{Pinchen Xie}
\affiliation{Program in Applied and Computational Mathematics, Princeton University, Princeton, NJ 08544, USA}
\author{Yixiao Chen}
\affiliation{Program in Applied and Computational Mathematics, Princeton University, Princeton, NJ 08544, USA}
\author{Pablo M. Piaggi}
\affiliation{CIC nanoGUNE BRTA, Tolosa Hiribidea 76, 20018 Donostia-San Sebasti\'an, Spain}
\affiliation{Ikerbasque, Basque Foundation for Science, 48013 Bilbao, Spain}
\author{Roberto Car}
\email{rcar@princeton.edu}
\affiliation{Department of Chemistry, Princeton University, Princeton, NJ 08544, USA}
\affiliation{Program in Applied and Computational Mathematics, Princeton University, Princeton, NJ 08544, USA}
\affiliation{Department of Physics, Princeton University, Princeton, NJ 08544, USA}
\affiliation{Princeton Institute for the Science and Technology of Materials, Princeton University, Princeton, NJ 08544, USA}

\begin{document}

\begin{abstract}
Accurate predictions of machine learning interatomic potentials (MLIPs) rely on training datasets covering relevant thermodynamic conditions and phases. Using water as an example, we show that insufficiently sampled datasets of MLIPs lead to wrong predictions of densities, melting temperatures, and nuclear quantum effects. In contrast, MLIPs trained on comprehensive datasets with the same electronic structure method yield consistent property predictions agnostic of MLIP architecture, with densities in quantitative agreement with reference calculations.
\end{abstract}

Machine learning interatomic potentials (MLIPs) enable calculations of thermodynamic properties that would be prohibitively expensive with direct \emph{ab initio} molecular dynamics (AIMD). For liquid water and ice, MLIPs have enabled studies of phase equilibria and phase diagrams.\cite{cheng_ab_2019,reinhardt_quantum-mechanical_2021,piaggi_phase_2021,zhang_phase_2021} In these calculations, the MLIP is a surrogate for an underlying electronic structure method. Different MLIP architectures representing the same density functional theory (DFT) settings should therefore predict consistent properties. In practice, however, MLIPs are usually trained on different datasets, and properties governed by small energy differences can vary significantly even when the same electronic structure settings are used. An example is the melting temperature ($\tm$) of ice: values obtained with different revPBE0-D3-based MLIPs differ by more than 10 K.\cite{cheng_ab_2019, chen_thermodynamics_2024} These discrepancies can arise from incomplete datasets: a dataset that does not cover the phases and thermodynamic conditions encountered in the target simulation can produce large generalization errors and unreliable property predictions. 

Water is an important yet challenging system for first-principles simulations, as demonstrated by the following two examples. The first is the $\sim$8\% density increase from ice Ih to liquid water, which underlies the familiar density anomaly of water. In contrast, AIMD simulations with the PBE functional incorrectly yield ice denser than liquid water\cite{chen_ab_2017}, giving $\Delta\rho=\rho_{\mathrm{liq}}-\rho_{\mathrm{ice}}=-0.086\pm0.021$ g/cm$^3$, where $\rho_{\mathrm{liq}}$ and $\rho_{\mathrm{ice}}$ denote the mass densities of liquid water and ice Ih, respectively. The second is $\tm$ and its shift due to nuclear quantum effects (NQEs). Experimentally, $\tm$ is 273.15 K for H$_2$O and 276.97 K for D$_2$O at 1 bar; this difference of 3.82 K shows that the stronger NQEs in H$_2$O lower its $\tm$ relative to D$_2$O.\cite{ramirez_quantum_2010-1} Both observables depend on a delicate balance between the thermodynamic properties of liquid water and ice.

In this Letter, we compare water and ice MLIPs constructed using the same revPBE0-D3 DFT setup to demonstrate that a converged dataset is essential for reliable thermodynamic predictions. The hybrid revPBE0-D3 functional was applied to AIMD simulations of liquid water in 2017\cite{marsalek_quantum_2017} and has since become widely used in MLIP-based simulations studying bulk water, ice, aqueous proton transport, and confined-water phase behavior.\cite{ruiz_pestana_quest_2018,cheng_ab_2019,schran_committee_2020,reinhardt_quantum-mechanical_2021,kapil_first-principles_2022,chen_thermodynamics_2024,montero_de_hijes_density_2024,berrens_nuclear_2024,gomez_neural-network-based_2024,shiga_computation_2025}

We construct a Deep Potential (DP) model of revPBE0-D3 water and ice using DP-GEN for active learning\cite{zhang_dp-gen_2020} and DeePMD-kit for fitting energies and forces with the \texttt{se\_e2\_a} descriptor and a 6~\AA{} cutoff radius.\cite{wang_deepmd-kit_2018} Starting from 100 liquid-water configurations, we iteratively train four DP models and explore liquid water at 270--350 K and ice Ih at 150--300 K in 64-molecule cells using classical molecular dynamics (MD) and 32-bead path integral molecular dynamics (PIMD) in the $NpT$ ensemble at 1 bar. For each explored configuration, DP-GEN evaluates the model deviation, defined as the maximum over atoms of the root-mean-square deviation of the four predicted force vectors from their mean across the four models. Configurations with model deviations of at least 0.2 but less than 0.35~eV/\AA{} are selected as candidates for revPBE0-D3 energy and force calculations; the labeled configurations are then added to the dataset before retraining. This cycle continues until more than 99.9\% of the explored configurations have a model deviation below 0.2~eV/\AA. The reference revPBE0-D3 calculations use CP2K with the same DFT settings as those used by Cheng et al.\ in Ref.~\citenum{cheng_ab_2019}, except that the plane-wave cutoff is increased from 400 to 800 Ry for improved convergence, following their later recommendation in Ref.~\citenum{monserrat_liquid_2020}. We refer to the resulting DP training dataset as SET2 and the dataset used to train the Behler--Parrinello neural network (BPNN) model of Ref.~\citenum{cheng_ab_2019} (hereafter BPNN1) as SET1. Model-training details are given in Section~S5 of the Supporting Information (SI).

Using our MLIP, we calculate several fundamental thermodynamic properties of liquid water and ice, including $\tm$, the density $\rho$ of both phases, and the temperature of maximum liquid-water density, $\tdm$. We obtain equilibrium densities from $NpT$ MD simulations at 1 bar and determine $\tdm$ from the density isobars shown in Figure~S10, with simulation details provided in Section~S7 of the SI. The classical melting temperature, $\tm^{\mathrm{cl}}$, is located where the ice and liquid chemical potentials are equal, with these chemical potentials calculated by thermodynamic integration (TI)\cite{zhang_phase_2021} using the \texttt{dpti} package\cite{yuan_dpti_2026}. We then add NQEs to the chemical potentials and determine $\tm^{\mathrm{qu}}$ through mass thermodynamic integration (MTI) using PIMD in the $NpT$ ensemble at 1 bar and a set of temperatures around $\tm^{\mathrm{cl}}$, with the \texttt{fix pimd/langevin} module of LAMMPS.\cite{thompson_lammps_2022} MTI integrates from the quantum limit with physical atomic masses to the classical limit of infinite masses. In our simulations, each physical atomic mass $m$ is scaled to $m/y^2$, where the dimensionless parameter $y$ varies from the classical infinite-mass limit ($y\to0$) to the physical masses ($y=1$). The integrand is
\begin{equation}
\Delta g_{\mathrm{ice}-\mathrm{liq}}(y)
=\frac{2}{y}\left[\langle K_{\mathrm{ice}}(y)\rangle-\langle K_{\mathrm{liq}}(y)\rangle\right],
\label{eq:mti-integrand}
\end{equation}
where $\langle K_{\mathrm{ice}}(y)\rangle$ and $\langle K_{\mathrm{liq}}(y)\rangle$ are the ensemble-averaged nuclear kinetic energies per water molecule in ice and liquid water at the scaled masses. We evaluate the MTI integrand at six values, $y=0.1$, 0.2, 0.3, 0.4, 0.6, and 1.0. At the classical endpoint, $\Delta g_{\mathrm{ice}-\mathrm{liq}}(0)=0$ by definition. Integrating this quantity from $y=0$ to $1$ gives the NQE correction to the ice--liquid chemical-potential difference. Adding this correction to the classical difference and locating its zero yields the quantum melting temperature, $\tm^{\mathrm{qu}}$, and hence the NQE-induced shift, $\Delta\tm=\tm^{\mathrm{qu}}-\tm^{\mathrm{cl}}$. Classical free-energy calculations and MTI details are given in Sections~S8 and~S9, respectively; the DP chemical-potential curves are shown in Figure~S11.

Table~\ref{tab:properties} compares our results using the DP model with those of previously reported revPBE0-D3-based MLIPs and with experiment, revealing substantial discrepancies among the models. BPNN1 yields $\tm^{\mathrm{cl}}=275$ K and an NQE-induced shift of $\Delta\tm=-8$ K, whereas our DP model gives $\tm^{\mathrm{cl}}=296$ K and an NQE-induced shift of $\Delta\tm=+7$ K. The neuroevolution potential (NEP) model trained on the same revPBE0-D3 DFT and reported in Ref.~\citenum{chen_thermodynamics_2024} yields a classical liquid density of 1.001 g/cm$^3$ at 298.15 K, while BPNN1 and our DP show 0.94 and 0.8910 g/cm$^3$ at 300 K, respectively. These discrepancies are much larger than expected for models intended to represent the same underlying level of theory.

\begin{table*}
\caption{Selected experimental and simulated properties of water and ice at 1 bar. $\tdm$ is the temperature of maximum liquid-water density. The superscripts $\mathrm{cl}$ and $\mathrm{qu}$ denote classical and quantum nuclei, respectively; experimental densities are those of physical water. Numbers in parentheses indicate statistical uncertainties in the last quoted digits. Densities and $\Delta\rho=\rho_{\mathrm{liq}}-\rho_{\mathrm{ice}}$ are reported in g/cm$^3$. The densities are reported at $T=300$ K for all simulations and $T=273$ K for experiment, unless otherwise noted. Gray-shaded rows report literature values; unshaded data rows report calculations from this work.}
\label{tab:properties}
{\setlength{\tabcolsep}{4pt}
\begin{tabular*}{\textwidth}{@{\extracolsep{\fill}}cccccccc}
\toprule
Source & $\tm^{\mathrm{cl}}$ [K] & $\tm^{\mathrm{qu}}$ [K] & $\Delta\tm$ [K] & $\rho_{\mathrm{liq}}^{\mathrm{cl}}$ & $\rho_{\mathrm{ice}}^{\mathrm{cl}}$ & $\Delta\rho$ & $\tdm$ [K] \\
\midrule
\rowcolor{literaturegray}
Experiment\cite{ceriotti_nuclear_2016} & $>277.64$\textsuperscript{a} & 273.15 & $<-4.49$ & 0.9998 & 0.917 & 0.083 & 277.13 \\
AIMD, this work & -- & -- & -- & 0.925(20) & 0.887(15) & 0.038(25) & -- \\
\rowcolor{literaturegray}
BPNN1, Ref.~\citenum{cheng_ab_2019} & 275(2) & 267(2) & $-8$ & 0.94 & 0.88 & 0.06 & $\approx 277$ \\
\rowcolor{literaturegray}
NEP, Ref.~\citenum{chen_thermodynamics_2024} & 290.5(5) & -- & -- & 1.001(2)\textsuperscript{b} & 0.915(2)\textsuperscript{b} & -- & 292.2(4) \\
DP, this work & 296(1) & 303(1) & $+7$ & 0.8910(6) & 0.8719(1) & 0.0191(6) & 285 \\
\bottomrule
\end{tabular*}}
\par\smallskip
\parbox{\textwidth}{\footnotesize\textsuperscript{a}Although classical water does not exist experimentally, its melting temperature is inferred to exceed 277.64 K from the isotope trend: D$_2$O and T$_2$O melt at 276.97 and 277.64 K, respectively.\cite{ceriotti_nuclear_2016}\par
\textsuperscript{b}The NEP liquid-water density is reported at 298.15 K and 1.01325 bar; the ice-Ih density is reported at 250 K and 1 bar.\cite{chen_thermodynamics_2024}}
\end{table*}

To determine which MLIP predictions reproduce the behavior of the underlying functional, we performed revPBE0-D3 AIMD simulations of liquid water and ice Ih at 300 K and 1 bar. The $NpT$ AIMD simulations yield densities of $0.925\pm0.020$ and $0.887\pm0.015$ g/cm$^3$ for liquid water and ice Ih, respectively. An independent liquid-water equation-of-state analysis from $NVT$ AIMD simulations gives $0.926\pm0.007$ g/cm$^3$, as shown in Figure~S3, validating the convergence of the $NpT$ AIMD simulations. Details of the AIMD calculations are reported in Section~S1 of the SI. BPNN1 overestimates the AIMD liquid-water density by 1.6\%, whereas DP underestimates it by 3.5\%; both models closely reproduce the AIMD ice density, yielding acceptable overall agreement with AIMD for the densities of both phases. Moreover, Figure~S3 shows that DP systematically overestimates the average internal pressure of liquid water at fixed density relative to AIMD, explaining its underestimated equilibrium density.

\begin{figure}[!t]
\centering
\includegraphics[width=0.5\linewidth]{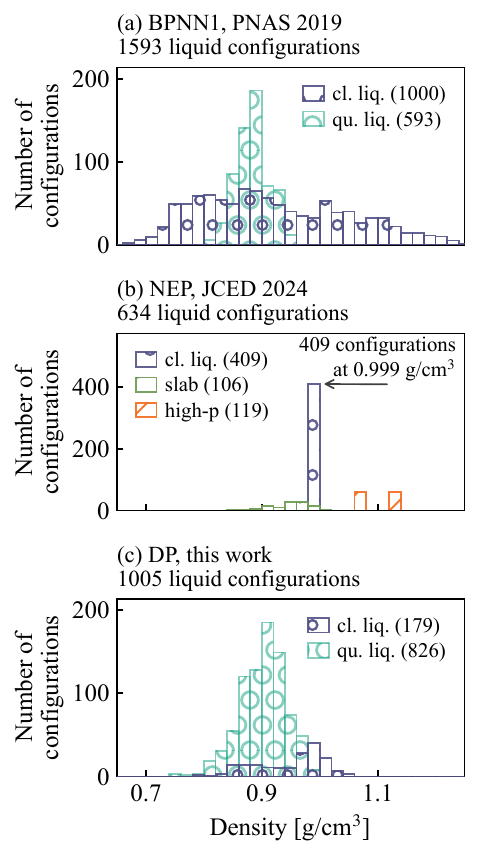}
\caption{Density distributions of liquid-water configurations in the revPBE0-D3 training datasets for (a) BPNN1, (b) NEP, and (c) DP, using a bin width of 0.02~g/cm$^3$. Ice configurations are omitted. Legend abbreviations cl., qu., liq., and high-p denote classical, quantum, liquid, and high-pressure configurations, respectively. All 409 classical bulk-water configurations in the NEP dataset have a density of 0.999~g/cm$^3$. For the water-slab configurations, liquid-region densities are calculated by averaging the densities within $\pm$10~\AA{} of the slab center; the simulation-cell volumes are identical across these configurations.}
\label{fig:density-coverage}
\end{figure}

However, despite its close agreement with experiment, the published NEP liquid density of 1.001 g/cm$^3$ exceeds the revPBE0-D3 AIMD value by 8.2\%. As shown in Figure~\ref{fig:density-coverage}, the NEP dataset contains 409 classical bulk-water configurations, all at a density of 0.999 g/cm$^3$, close to the experimental value. It also includes high-pressure water and water-slab configurations, but these environments differ from ambient-pressure bulk liquid water and do not directly sample its density fluctuations. This restricted bulk-liquid density sampling during training leads NEP to predict a density close to input data rather than the lower revPBE0-D3 AIMD value. Its systematic energy bias on independent $NpT$ AIMD configurations further demonstrates this mismatch with the underlying DFT reference, as shown in Figure~S7. Figure~S4 provides a configuration-by-configuration comparison of the density coverage of the three datasets. By contrast, as shown in Figure~\ref{fig:density-coverage}, the datasets used to train BPNN1 and DP, SET1 and SET2 respectively, cover the density range sampled by the revPBE0-D3 $NpT$ AIMD trajectory of liquid water given in Figure~S1, from approximately 0.85 to 1.05 g/cm$^3$. To test this interpretation, we trained an NEP on SET3, an expanded version of SET2 obtained through the active-learning procedure described below. At 300 K and 1 bar, this model yields classical liquid-water and ice densities of $0.9014\pm0.0006$ and $0.8886\pm0.0004$ g/cm$^3$, respectively (Section~S6.2 of the SI). These values are in acceptable agreement with the corresponding 64-molecule DP densities of $0.900\pm0.003$ and $0.874\pm0.002$ g/cm$^3$ (Section~S1.2 of the SI). These results support the importance of sampling the density fluctuations of the target $NpT$ ensemble in the training data. Density and energy distribution analyses are detailed in Section~S2.

\begin{figure}[!htb]
\centering
\includegraphics[width=0.5\linewidth]{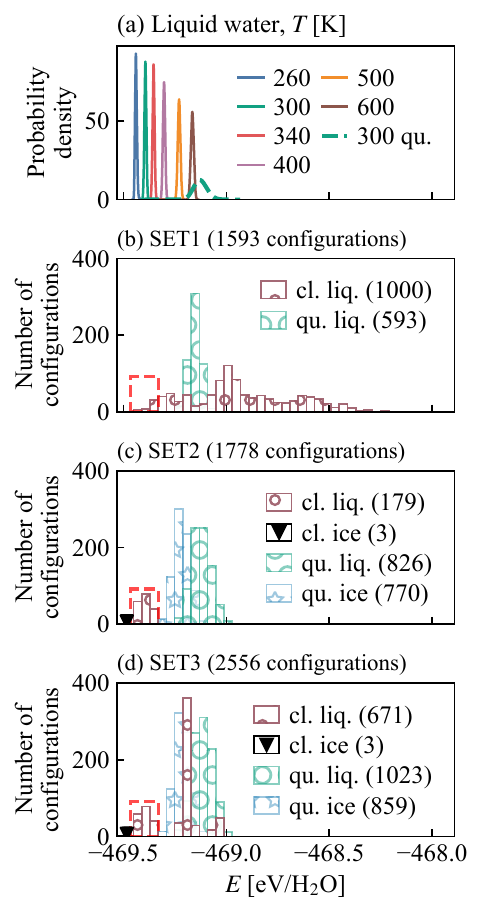}
\caption{Potential-energy distributions of liquid water and the revPBE0-D3 training configurations. $E$ denotes the potential energy per water molecule. (a) Classical liquid-water distributions from 260 to 600 K (solid lines) and the 300 K quantum distribution from 32-bead PIMD (dashed line). (b--d) Energy distributions of SET1, SET2, and the converged SET3 (2556 configurations), respectively. Legend abbreviations cl., qu., and liq. denote classical, quantum, and liquid, respectively. Red rectangles highlight classical liquid water configurations corresponding to room temperatures. SET1 contains only liquid-water configurations, whereas SET2 and SET3 include both liquid water and ice sampled from classical and path integral simulations.}
\label{fig:dataset}
\end{figure}

The most important qualitative difference between BPNN1 and DP is the sign of the NQE-induced melting-temperature shift, $\Delta\tm$. BPNN1 predicts a negative shift, consistent with experiment, whereas DP predicts a positive shift. We show that this difference also originates from their training data. As shown in Figure~\ref{fig:dataset}, SET1 for BPNN1 and SET2 for DP differ in two key respects. First, SET2 contains both liquid water and ice, whereas SET1 contains only liquid water; each phase included in either dataset is sampled with both classical and quantum nuclei. Second, our SET2 includes classical liquid-water configurations sampled around room temperature, covering the conditions of our subsequent simulations, whereas the classical liquid-water configurations in SET1 predominantly correspond to temperatures several hundred kelvin above room temperature. 

Unlike the density calculations above, determining $\tm^{\mathrm{qu}}$ by MTI requires long PIMD simulations at multiple mass values, making direct AIMD validation prohibitively expensive. We therefore test the DP prediction using the BPNN architecture. However, a BPNN model trained on SET2 (hereafter BPNN2) becomes unstable in extended MD and PIMD simulations, as shown in Figure~S5 and discussed in Section~S3. This instability shows that BPNN requires more training data than DP under these conditions. Starting from SET2, we perform additional BPNN active-learning iterations using the same force-deviation thresholds and convergence criterion as for DP. Once more than 99.9\% of the explored configurations have a model deviation below 0.2~eV/\AA, we obtain the 2556-configuration dataset SET3 and train BPNN3 on it. BPNN3 and DP thus represent two distinct architectures trained on comprehensive datasets covering liquid water and ice at 1 bar and around room temperature. Dataset sizes are summarized in Table~S1.

Figure~\ref{fig:dataset}(a) shows the equilibrium potential-energy distributions sampled with the DP model at several temperatures, while panels (b--d) show the distributions of the training datasets. The classical liquid-water distributions are narrow and overlap little even for temperatures separated by only 40 K. Thus, classical configurations sampled at high temperatures provide limited coverage of the energy range relevant to low-temperature simulations. The energy distribution of the liquid water configurations sampled in PIMD simulations is broader and extends to higher energies: at 300 K, it overlaps substantially with the classical distribution at 600 K. SET1 contains few classical liquid-water configurations in the room-temperature energy range highlighted by the red rectangles; most instead lie in the higher-energy ranges associated with substantially higher temperatures. It also lacks ice configurations. SET2 and SET3 cover the room-temperature liquid configurations and include both phases with classical and quantum nuclei. The consequences for model accuracy are evident in Figure~S6, which compares predictions against independent AIMD test sets containing 50 liquid-water and 50 ice configurations at 300 K and 1 bar, with test-set construction and validation details in Sections~S1.3 and S4. BPNN1 exhibits a systematic positive energy bias for liquid water and larger force errors, whereas BPNN3 has substantially smaller errors and performs comparably to DP. Together, the improved dataset coverage and independent test accuracy show that the discrepancies between BPNN1 and DP originate from the training data and are agnostic of the MLIP architecture.

\begin{figure}[!htb]
\centering
\includegraphics[width=0.5\linewidth]{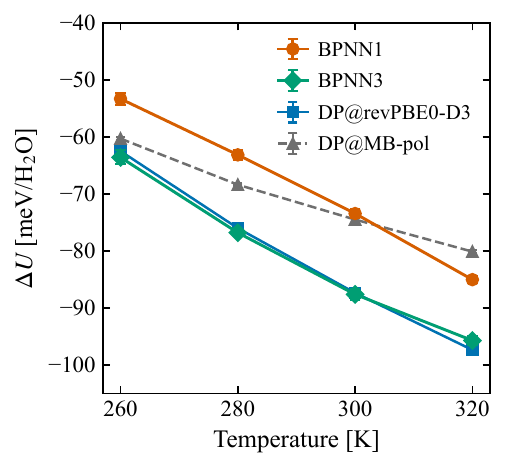}
\caption{Classical ice--liquid energy difference per water molecule, $\Delta U(T)=U_{\mathrm{ice}}-U_{\mathrm{liq}}$, at 1 bar for BPNN1, BPNN3, and DP trained on revPBE0-D3 data. DP@MB-pol is included for comparison.}
\label{fig:delta-u}
\end{figure}

The missing energy range in SET1 leads to systematic biases in BPNN1's energy predictions. BPNN1 overestimates the energies of ice and liquid water by different amounts, underestimating the magnitude of the classical ice--liquid energy difference, $\Delta U(T)=U_{\mathrm{ice}}-U_{\mathrm{liq}}$, as shown in Figure~\ref{fig:delta-u}. Here $U_{\mathrm{ice}}$ and $U_{\mathrm{liq}}$ are the mean potential energies per water molecule in ice Ih and liquid water, respectively, obtained from $NpT$ MD simulations at 1 bar using 64-molecule cells for BPNN1 and BPNN3 and 432-molecule cells for the DP references. We include DP@MB-pol, a DP model trained on the MB-pol potential with a reported classical melting temperature of 262.5 K, as a reference.\cite{bore_realistic_2023} Across 260--320 K, BPNN3 closely follows DP, whereas BPNN1 predicts a less negative $\Delta U$ and lies closer to DP@MB-pol. At 300 K, for example, $\Delta U$ is $-73.4\pm0.8$, $-87.6\pm0.9$, and $-87.4\pm0.3$ meV/H$_2$O for BPNN1, BPNN3, and DP, respectively. The less negative value means that BPNN1 provides weaker energetic stabilization of ice relative to liquid water, consistent with BPNN1's lower classical melting temperature and closer agreement with experiment.

\begin{figure}[!htb]
\centering
\includegraphics[width=\textwidth]{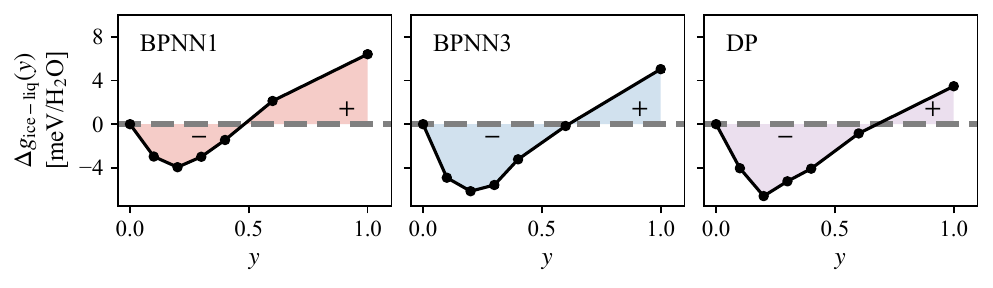}
\caption{MTI integrand $\Delta g_{\mathrm{ice}-\mathrm{liq}}(y)$ calculated at 290 K and 1 bar using BPNN1 trained on SET1, BPNN3 trained on the actively learned SET3, and DP trained on SET2. Positive and negative shaded areas denote contributions of opposite sign to the integral.}
\label{fig:mti-dataset}
\end{figure}

The discrepancy between BPNN1 and DP in the sign of the NQE-induced melting-temperature shift is also resolved by BPNN3, trained on the more comprehensive SET3. Figure~\ref{fig:mti-dataset} compares the MTI integrands defined in Eq.~\eqref{eq:mti-integrand} for BPNN1, BPNN3, and DP at 290 K and 1 bar. The bead counts and sampling lengths are given in Table~S2. For BPNN1, the positive contribution exceeds the negative contribution, giving an integral of $0.71\pm0.08$ meV/H$_2$O. This positive correction raises the ice chemical potential relative to that of the liquid and therefore lowers the melting temperature. For BPNN3 and DP, the negative contribution dominates, yielding integrals of $-1.19\pm0.08$ and $-1.76\pm0.02$ meV/H$_2$O, respectively. NQEs therefore stabilize ice relative to liquid water and raise the melting temperature for both models. BPNN3 thus recovers the positive sign of $\Delta\tm=+7$ K obtained with DP, as reported in Table~\ref{tab:properties}, using an independent architecture, whereas BPNN1 yields a negative shift ($\Delta\tm=-8$ K). Together with the classical energetic comparison in Figure~\ref{fig:delta-u}, these results demonstrate that converging the training dataset is essential to reproduce the revPBE0-D3 prediction, even when that prediction disagrees with experiment.

Training dataset coverage determines whether an MLIP faithfully reproduces the predictions of its underlying reference method. Using water as an example, we show that differences in training datasets lead to inconsistent predictions of density, relative stability of phases, and NQE-induced melting temperature shifts among MLIPs trained on the same DFT settings. Restricted liquid density sampling causes the NEP in Ref.~\citenum{chen_thermodynamics_2024} to reproduce a density close to experiment rather than the lower revPBE0-D3 AIMD value. For BPNN1 in Ref.~\citenum{cheng_ab_2019}, sparse room-temperature liquid coverage and the absence of ice configurations lead to an underestimated ice--liquid water energy difference, a lower melting temperature, and the opposite sign of the NQE-induced melting temperature shift compared to DP. BPNN3 trained on a comprehensive dataset recovers the DP energetics and the sign of the NQE-induced melting temperature shift. Reliable thermodynamic predictions from MLIPs require training data that cover the relevant phases and conditions.

\section*{Associated Content}
\subsection*{Data Availability Statement}
The models, training data, and input files required to reproduce this work are available in the GitHub repository at \url{https://github.com/Yi-FanLi/NQE-Ice-Tm/}.\cite{li_httpsgithubcomyi-fanlinqe-ice-tm_nodate}

\subsection*{Supporting Information}
AIMD reference calculations; training-set coverage, model training, and validation; MD and PIMD settings; classical free-energy calculations and mass thermodynamic integration (PDF)

\section*{Acknowledgments}
We thank Han Wang, Jinzhe Zeng, and Bingqing Cheng for useful discussions, and Jinggang Lan for providing the BPNN training input files. This work is supported by the U.S. Department of Energy through the Computational Chemical Sciences Center ``Chemistry in Solution and at Interfaces'' under Award No. DE-SC0019394 and ``Multiscale Chemistry and Physics in Materials and Fluids'' under Award No. DE-SC0026745.

\bibliography{clean}

@Article{	  bore_realistic_2023,
  title		= {Realistic phase diagram of water from “first
		  principles” data-driven quantum simulations},
  volume	= {14},
  copyright	= {2023 The Author(s)},
  issn		= {2041-1723},
  url		= {https://www.nature.com/articles/s41467-023-38855-1},
  doi		= {10.1038/s41467-023-38855-1},
  number	= {1},
  urldate	= {2024-06-19},
  journal	= {Nature Communications},
  author	= {Bore, Sigbjørn Løland and Paesani, Francesco},
  month		= jun,
  year		= {2023},
  pages		= {3349}
}

@Article{berrens_nuclear_2024,
  title = {Nuclear Quantum Effects on the Electronic Structure of Water and Ice},
  author = {Berrens, Margaret L. and Kundu, Arpan and Calegari Andrade, Marcos F. and Pham, Tuan Anh and Galli, Giulia and Donadio, Davide},
  journal = {The Journal of Physical Chemistry Letters},
  year = {2024},
  volume = {15},
  number = {26},
  pages = {6818--6825},
  doi = {10.1021/acs.jpclett.4c01315},
  url = {https://doi.org/10.1021/acs.jpclett.4c01315}
}

@Article{	  chen_thermodynamics_2024,
  title		= {Thermodynamics of {Water} and {Ice} from a {Fast} and
		  {Scalable} {First}-{Principles} {Neuroevolution}
		  {Potential}},
  volume	= {69},
  issn		= {0021-9568},
  url		= {https://doi.org/10.1021/acs.jced.3c00561},
  doi		= {10.1021/acs.jced.3c00561},
  number	= {1},
  urldate	= {2024-04-09},
  journal	= {Journal of Chemical \& Engineering Data},
  author	= {Chen, Zekun and Berrens, Margaret L. and Chan, Kam-Tung
		  and Fan, Zheyong and Donadio, Davide},
  month		= jan,
  year		= {2024},
  pages		= {128--140}
}

@Article{	  wang_deepmd-kit_2018,
  title		= {{DeePMD}-kit: {A} deep learning package for many-body
		  potential energy representation and molecular dynamics},
  volume	= {228},
  issn		= {0010-4655},
  shorttitle	= {{DeePMD}-kit},
  url		= {https://www.sciencedirect.com/science/article/pii/S0010465518300882},
  doi		= {10.1016/j.cpc.2018.03.016},
  urldate	= {2023-01-07},
  journal	= {Computer Physics Communications},
  author	= {Wang, Han and Zhang, Linfeng and Han, Jiequn and E,
		  Weinan},
  month		= jul,
  year		= {2018},
  pages		= {178--184}
}

@Article{	  zhang_phase_2021,
  title		= {Phase {Diagram} of a {Deep} {Potential} {Water} {Model}},
  volume	= {126},
  url		= {https://link.aps.org/doi/10.1103/PhysRevLett.126.236001},
  doi		= {10.1103/PhysRevLett.126.236001},
  number	= {23},
  urldate	= {2023-01-07},
  journal	= {Physical Review Letters},
  author	= {Zhang, Linfeng and Wang, Han and Car, Roberto and E,
		  Weinan},
  month		= jun,
  year		= {2021},
  pages		= {236001}
}

@Article{	  zhang_dp-gen_2020,
  title		= {{DP}-{GEN}: {A} concurrent learning platform for the
		  generation of reliable deep learning based potential energy
		  models},
  volume	= {253},
  issn		= {0010-4655},
  shorttitle	= {{DP}-{GEN}},
  url		= {https://www.sciencedirect.com/science/article/pii/S001046552030045X},
  doi		= {10.1016/j.cpc.2020.107206},
  urldate	= {2023-01-07},
  journal	= {Computer Physics Communications},
  author	= {Zhang, Yuzhi and Wang, Haidi and Chen, Weijie and Zeng,
		  Jinzhe and Zhang, Linfeng and Wang, Han and E, Weinan},
  month		= aug,
  year		= {2020},
  pages		= {107206}
}

@Article{	  piaggi_phase_2021,
  title		= {Phase {Equilibrium} of {Water} with {Hexagonal} and
		  {Cubic} {Ice} {Using} the {SCAN} {Functional}},
  volume	= {17},
  issn		= {1549-9618, 1549-9626},
  url		= {https://pubs.acs.org/doi/10.1021/acs.jctc.1c00041},
  doi		= {10.1021/acs.jctc.1c00041},
  number	= {5},
  urldate	= {2023-01-05},
  journal	= {Journal of Chemical Theory and Computation},
  author	= {Piaggi, Pablo M. and Panagiotopoulos, Athanassios Z. and
		  Debenedetti, Pablo G. and Car, Roberto},
  month		= may,
  year		= {2021},
  pages		= {3065--3077}
}

@Article{	  thompson_lammps_2022,
  title		= {{LAMMPS} - a flexible simulation tool for particle-based
		  materials modeling at the atomic, meso, and continuum
		  scales},
  volume	= {271},
  issn		= {0010-4655},
  url		= {https://www.sciencedirect.com/science/article/pii/S0010465521002836},
  doi		= {10.1016/j.cpc.2021.108171},
  urldate	= {2023-01-05},
  journal	= {Computer Physics Communications},
  author	= {Thompson, Aidan P. and Aktulga, H. Metin and Berger,
		  Richard and Bolintineanu, Dan S. and Brown, W. Michael and
		  Crozier, Paul S. and in 't Veld, Pieter J. and Kohlmeyer,
		  Axel and Moore, Stan G. and Nguyen, Trung Dac and Shan, Ray
		  and Stevens, Mark J. and Tranchida, Julien and Trott,
		  Christian and Plimpton, Steven J.},
  month		= feb,
  year		= {2022},
  pages		= {108171}
}

@Article{	  cheng_ab_2019,
  title		= {Ab initio thermodynamics of liquid and solid water},
  volume	= {116},
  url		= {https://www.pnas.org/doi/abs/10.1073/pnas.1815117116},
  doi		= {10.1073/pnas.1815117116},
  number	= {4},
  urldate	= {2022-12-01},
  journal	= {Proceedings of the National Academy of Sciences},
  author	= {Cheng, Bingqing and Engel, Edgar A. and Behler, Jörg and
		  Dellago, Christoph and Ceriotti, Michele},
  month		= jan,
  year		= {2019},
  pages		= {1110--1115}
}

@Article{	  reinhardt_quantum-mechanical_2021,
  title		= {Quantum-mechanical exploration of the phase diagram of
		  water},
  volume	= {12},
  copyright	= {2021 The Author(s)},
  issn		= {2041-1723},
  url		= {https://www.nature.com/articles/s41467-020-20821-w},
  doi		= {10.1038/s41467-020-20821-w},
  number	= {1},
  urldate	= {2022-12-01},
  journal	= {Nature Communications},
  author	= {Reinhardt, Aleks and Cheng, Bingqing},
  month		= jan,
  year		= {2021},
  pages		= {588}
}

@Article{	  marsalek_quantum_2017,
  title		= {Quantum {Dynamics} and {Spectroscopy} of {Ab} {Initio}
		  {Liquid} {Water}: {The} {Interplay} of {Nuclear} and
		  {Electronic} {Quantum} {Effects}},
  volume	= {8},
  shorttitle	= {Quantum {Dynamics} and {Spectroscopy} of {Ab} {Initio}
		  {Liquid} {Water}},
  url		= {https://doi.org/10.1021/acs.jpclett.7b00391},
  doi		= {10.1021/acs.jpclett.7b00391},
  number	= {7},
  urldate	= {2022-11-29},
  journal	= {The Journal of Physical Chemistry Letters},
  author	= {Marsalek, Ondrej and Markland, Thomas E.},
  month		= apr,
  year		= {2017},
  pages		= {1545--1551}
}

@Article{	  ruiz_pestana_quest_2018,
  title		= {The {Quest} for {Accurate} {Liquid} {Water} {Properties}
		  from {First} {Principles}},
  volume	= {9},
  issn		= {1948-7185, 1948-7185},
  url		= {https://pubs.acs.org/doi/10.1021/acs.jpclett.8b02400},
  doi		= {10.1021/acs.jpclett.8b02400},
  number	= {17},
  urldate	= {2022-11-29},
  journal	= {The Journal of Physical Chemistry Letters},
  author	= {Ruiz Pestana, Luis and Marsalek, Ondrej and Markland,
		  Thomas E. and Head-Gordon, Teresa},
  month		= sep,
  year		= {2018},
  pages		= {5009--5016}
}

@Article{	  ceriotti_nuclear_2016,
  title		= {Nuclear {Quantum} {Effects} in {Water} and {Aqueous}
		  {Systems}: {Experiment}, {Theory}, and {Current}
		  {Challenges}},
  volume	= {116},
  issn		= {0009-2665, 1520-6890},
  shorttitle	= {Nuclear {Quantum} {Effects} in {Water} and {Aqueous}
		  {Systems}},
  url		= {https://pubs.acs.org/doi/10.1021/acs.chemrev.5b00674},
  doi		= {10.1021/acs.chemrev.5b00674},
  number	= {13},
  urldate	= {2022-11-28},
  journal	= {Chemical Reviews},
  author	= {Ceriotti, Michele and Fang, Wei and Kusalik, Peter G. and
		  McKenzie, Ross H. and Michaelides, Angelos and Morales,
		  Miguel A. and Markland, Thomas E.},
  month		= jul,
  year		= {2016},
  pages		= {7529--7550}
}

@Article{	  chen_ab_2017,
  title		= {Ab initio theory and modeling of water},
  volume	= {114},
  url		= {https://www.pnas.org/doi/abs/10.1073/pnas.1712499114},
  doi		= {10.1073/pnas.1712499114},
  number	= {41},
  urldate	= {2022-11-28},
  journal	= {Proceedings of the National Academy of Sciences},
  author	= {Chen, Mohan and Ko, Hsin-Yu and Remsing, Richard C. and
		  Calegari Andrade, Marcos F. and Santra, Biswajit and Sun,
		  Zhaoru and Selloni, Annabella and Car, Roberto and Klein,
		  Michael L. and Perdew, John P. and Wu, Xifan},
  month		= oct,
  year		= {2017},
  pages		= {10846--10851}
}

@Article{	  ramirez_quantum_2010-1,
  title		= {Quantum path integral simulation of isotope effects in the
		  melting temperature of ice {Ih}},
  volume	= {133},
  issn		= {0021-9606},
  url		= {https://doi.org/10.1063/1.3503764},
  doi		= {10.1063/1.3503764},
  number	= {14},
  urldate	= {2024-03-03},
  journal	= {The Journal of Chemical Physics},
  author	= {Ramírez, R. and Herrero, C. P.},
  month		= oct,
  year		= {2010},
  pages		= {144511}
}

@Article{	  monserrat_liquid_2020,
  title		= {Liquid water contains the building blocks of diverse ice
		  phases},
  volume	= {11},
  copyright	= {2020 The Author(s)},
  issn		= {2041-1723},
  url		= {https://www.nature.com/articles/s41467-020-19606-y},
  doi		= {10.1038/s41467-020-19606-y},
  number	= {1},
  urldate	= {2024-02-16},
  journal	= {Nature Communications},
  author	= {Monserrat, Bartomeu and Brandenburg, Jan Gerit and Engel,
		  Edgar A. and Cheng, Bingqing},
  month		= nov,
  year		= {2020},
  pages		= {5757}
}

@Article{	  kapil_first-principles_2022,
  title		= {The first-principles phase diagram of monolayer
		  nanoconfined water},
  volume	= {609},
  copyright	= {2022 The Author(s), under exclusive licence to Springer
		  Nature Limited},
  issn		= {1476-4687},
  url		= {https://www.nature.com/articles/s41586-022-05036-x},
  doi		= {10.1038/s41586-022-05036-x},
  number	= {7927},
  urldate	= {2023-11-03},
  journal	= {Nature},
  author	= {Kapil, Venkat and Schran, Christoph and Zen, Andrea and
		  Chen, Ji and Pickard, Chris J. and Michaelides, Angelos},
  month		= sep,
  year		= {2022},
  pages		= {512--516}
}

@Article{	  hutter_cp2k_2014,
  title		= {cp2k: atomistic simulations of condensed matter systems},
  volume	= {4},
  copyright	= {© 2013 John Wiley \& Sons, Ltd.},
  issn		= {1759-0884},
  shorttitle	= {cp2k},
  url		= {https://onlinelibrary.wiley.com/doi/abs/10.1002/wcms.1159},
  doi		= {10.1002/wcms.1159},
  number	= {1},
  urldate	= {2024-07-16},
  journal	= {WIREs Computational Molecular Science},
  author	= {Hutter, Jürg and Iannuzzi, Marcella and Schiffmann,
		  Florian and VandeVondele, Joost},
  year		= {2014},
  pages		= {15--25}
}

@Article{	  gaiduk_density_2015,
  title		= {Density and {Compressibility} of {Liquid} {Water} and
		  {Ice} from {First}-{Principles} {Simulations} with {Hybrid}
		  {Functionals}},
  volume	= {6},
  url		= {https://doi.org/10.1021/acs.jpclett.5b00901},
  doi		= {10.1021/acs.jpclett.5b00901},
  number	= {15},
  urldate	= {2024-11-12},
  journal	= {The Journal of Physical Chemistry Letters},
  author	= {Gaiduk, Alex P. and Gygi, François and Galli, Giulia},
  month		= aug,
  year		= {2015},
  pages		= {2902--2908}
}

@Article{	  fan_neuroevolution_2021,
  title		= {Neuroevolution machine learning potentials: {Combining}
		  high accuracy and low cost in atomistic simulations and
		  application to heat transport},
  volume	= {104},
  shorttitle	= {Neuroevolution machine learning potentials},
  url		= {https://link.aps.org/doi/10.1103/PhysRevB.104.104309},
  doi		= {10.1103/PhysRevB.104.104309},
  number	= {10},
  urldate	= {2024-11-22},
  journal	= {Physical Review B},
  author	= {Fan, Zheyong and Zeng, Zezhu and Zhang, Cunzhi and Wang,
		  Yanzhou and Song, Keke and Dong, Haikuan and Chen, Yue and
		  Ala-Nissila, Tapio},
  month		= sep,
  year		= {2021},
  pages		= {104309}
}

@Article{	  schran_committee_2020,
  title		= {Committee neural network potentials control generalization
		  errors and enable active learning},
  volume	= {153},
  issn		= {0021-9606},
  url		= {https://doi.org/10.1063/5.0016004},
  doi		= {10.1063/5.0016004},
  number	= {10},
  urldate	= {2024-11-22},
  journal	= {The Journal of Chemical Physics},
  author	= {Schran, Christoph and Brezina, Krystof and Marsalek,
		  Ondrej},
  month		= sep,
  year		= {2020},
  pages		= {104105}
}

@Article{	  montero_de_hijes_density_2024,
  title		= {Density isobar of water and melting temperature of ice:
		  {Assessing} common density functionals},
  volume	= {161},
  issn		= {0021-9606},
  shorttitle	= {Density isobar of water and melting temperature of ice},
  url		= {https://doi.org/10.1063/5.0227514},
  doi		= {10.1063/5.0227514},
  number	= {13},
  urldate	= {2024-11-22},
  journal	= {The Journal of Chemical Physics},
  author	= {Montero de Hijes, Pablo and Dellago, Christoph and
		  Jinnouchi, Ryosuke and Kresse, Georg},
  month		= oct,
  year		= {2024},
  pages		= {131102}
}

@Article{	  fan_gpumd_2022,
  title		= {{GPUMD}: {A} package for constructing accurate
		  machine-learned potentials and performing highly efficient
		  atomistic simulations},
  volume	= {157},
  issn		= {0021-9606},
  shorttitle	= {{GPUMD}},
  url		= {https://doi.org/10.1063/5.0106617},
  doi		= {10.1063/5.0106617},
  number	= {11},
  urldate	= {2025-02-17},
  journal	= {The Journal of Chemical Physics},
  author	= {Fan, Zheyong and Wang, Yanzhou and Ying, Penghua and Song,
		  Keke and Wang, Junjie and Wang, Yong and Zeng, Zezhu and
		  Xu, Ke and Lindgren, Eric and Rahm, J. Magnus and Gabourie,
		  Alexander J. and Liu, Jiahui and Dong, Haikuan and Wu,
		  Jianyang and Chen, Yue and Zhong, Zheng and Sun, Jian and
		  Erhart, Paul and Su, Yanjing and Ala-Nissila, Tapio},
  month		= sep,
  year		= {2022},
  pages		= {114801}
}

@Article{	  gomez_neural-network-based_2024,
  title		= {Neural-network-based molecular dynamics simulations reveal
		  that proton transport in water is doubly gated by
		  sequential hydrogen-bond exchange},
  volume	= {16},
  copyright	= {2024 The Author(s), under exclusive licence to Springer
		  Nature Limited},
  issn		= {1755-4349},
  url		= {https://www.nature.com/articles/s41557-024-01593-y},
  doi		= {10.1038/s41557-024-01593-y},
  number	= {11},
  urldate	= {2025-09-23},
  journal	= {Nature Chemistry},
  author	= {Gomez, Axel and Thompson, Ward H. and Laage, Damien},
  month		= nov,
  year		= {2024},
  pages		= {1838--1844}
}

@Article{	  shiga_computation_2025,
  title		= {Computation of the heat capacity of water from first
		  principles},
  volume	= {163},
  issn		= {0021-9606},
  url		= {https://doi.org/10.1063/5.0285698},
  doi		= {10.1063/5.0285698},
  number	= {13},
  urldate	= {2025-10-16},
  journal	= {The Journal of Chemical Physics},
  author	= {Shiga, Motoyuki and Elsner, Jan and Behler, Jörg and
		  Thomsen, Bo},
  month		= oct,
  year		= {2025},
  pages		= {134119}
}

@Misc{		  li_httpsgithubcomyi-fanlinqe-ice-tm_nodate,
  title		= {https://github.com/{Yi}-{FanLi}/{NQE}-{Ice}-{Tm}},
  shorttitle	= {Yi-{FanLi}/{NQE}-{Ice}-{Tm}},
  url		= {https://github.com/Yi-FanLi/NQE-Ice-Tm},
  urldate	= {2025-12-09},
  journal	= {GitHub},
  author	= {Li, Yifan}
}

@Article{	  hare_density_1987,
  title		= {The density of supercooled water. {II}. {Bulk} samples
		  cooled to the homogeneous nucleation limit},
  volume	= {87},
  issn		= {0021-9606},
  url		= {https://doi.org/10.1063/1.453710},
  doi		= {10.1063/1.453710},
  number	= {8},
  urldate	= {2026-10-09},
  journal	= {The Journal of Chemical Physics},
  author	= {Hare, D. E. and Sorensen, C. M.},
  month		= oct,
  year		= {1987},
  pages		= {4840--4845}
}

@misc{yuan_dpti_2026,
  title = {{dpti}: An Automated Thermodynamic Integration Workflow for Phase Diagram Calculations with Machine Learning Interatomic Potentials},
  author = {Yuan, Fengbo and Zhong, Xin and Zheng, Donghao and Zeng, Jinzhe and Zhang, Linfeng and Wang, Han and Li, Yifan},
  year = {2026},
  doi = {10.48550/arXiv.2607.05015},
  url = {https://arxiv.org/abs/2607.05015},
  note = {Preprint, arXiv:2607.05015}
}

\renewcommand{\tocentryname}{For Table of Contents Only}
\begin{tocentry}
\includegraphics[width=3.25in,height=1.75in]{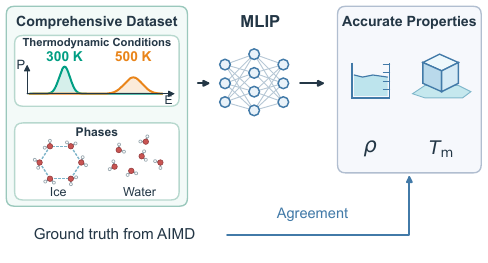}
\end{tocentry}

\end{document}


\section{AIMD Reference Calculations}\label{sec:aimd}

\subsection{Computational Details}

Reference revPBE0-D3 calculations were performed with CP2K version 2022.1.\cite{hutter_cp2k_2014} The auxiliary density matrix method (ADMM) was used to accelerate the exact-exchange calculation, and dual-space Goedecker--Teter--Hutter pseudopotentials were used to represent the core electrons. The settings followed previous revPBE0-D3 simulations of water,\cite{cheng_ab_2019} except that the plane-wave cutoff was increased to 800 Ry for improved convergence. The zero-damping form of the D3 dispersion correction was used. The orbital and auxiliary basis sets were TZV2P-GTH and cpFIT3, respectively; the SCF convergence threshold was $5\times10^{-7}$ Hartree.

Born--Oppenheimer AIMD simulations of liquid water and ice Ih contained 64 H$_2$O molecules and used a 2 fs timestep. Simulations were performed in the $NpT$ ensemble at 300 K and 1 bar. The masses of both O and H were set to 16 atomic mass units to permit the 2 fs timestep; this mass choice does not change equilibrium configurational averages such as the density. Temperature was controlled with a Nos\'e--Hoover chain thermostat of chain length three and damping time 160 fs. Pressure was controlled with an isotropic barostat for liquid water and a fully flexible cell barostat for ice Ih, both with a damping time of 800 fs.

\subsection{Density Validation}

\begin{figure}[!htb]
\centering
\includegraphics[width=0.70\textwidth]{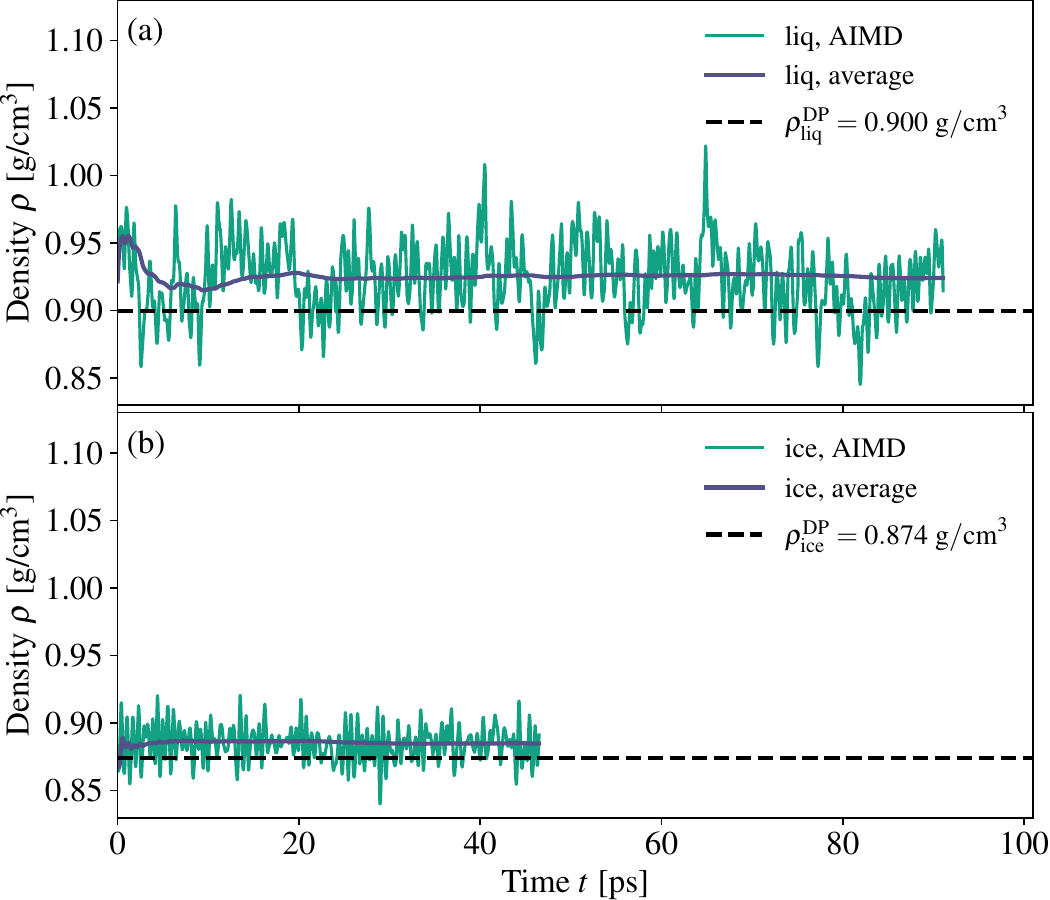}
\caption{Density trajectories of (a) liquid water and (b) ice Ih from revPBE0-D3 AIMD simulations in the $NpT$ ensemble at 300 K and 1 bar. Green curves show the instantaneous density, purple curves show the accumulated average, and black dashed lines show the average density from 64-molecule DP simulations under the same thermodynamic conditions.}
\label{figS:aimd-density-npt}
\end{figure}

Figure~\ref{figS:aimd-density-npt} shows the density trajectories and their accumulated averages. The liquid trace concatenates five runs after excluding the first 15,000 saved frames (30 ps) of each; its horizontal axis is accumulated sampling time across runs. The ice trace uses a single run. The direct $NpT$ AIMD estimates of density $\rho$ are $0.925\pm0.020$ g/cm$^3$ for liquid water and $0.887\pm0.015$ g/cm$^3$ for ice Ih. Corresponding 64-molecule DP simulations give $0.900\pm0.003$ and $0.874\pm0.002$ g/cm$^3$, respectively. Thus, DP reproduces the low-density revPBE0-D3 trend, although its mean densities are systematically lower than the AIMD estimates.

\begin{figure}[!htb]
\centering
\includegraphics[width=0.70\textwidth]{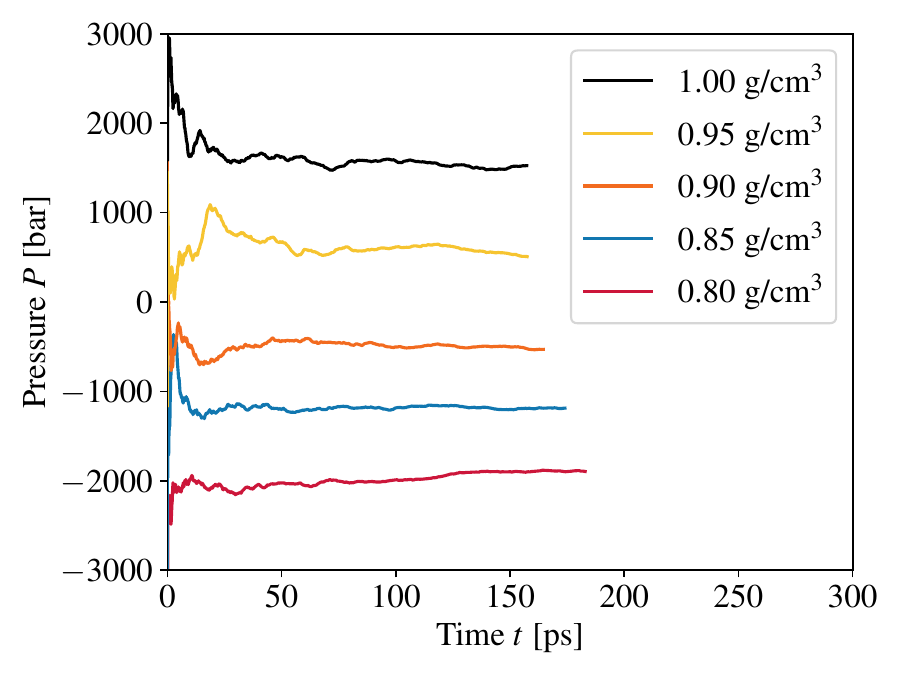}
\caption{Accumulated averages of the internal pressure from revPBE0-D3 liquid-water AIMD simulations in the $NVT$ ensemble at 300 K and the indicated densities.}
\label{figS:aimd-pressure-trajectories}
\end{figure}

\begin{figure}[!htb]
\centering
\includegraphics[width=0.62\textwidth]{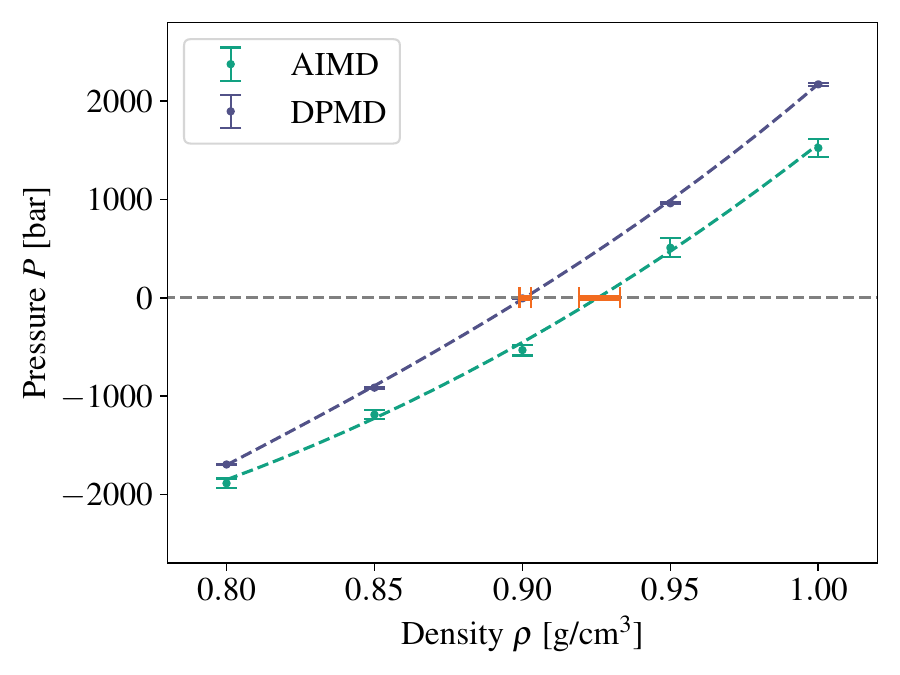}
\caption{Average internal pressure as a function of density from revPBE0-D3 AIMD and DP simulations of liquid water at 300 K in the $NVT$ ensemble. Cubic polynomials are fitted to the two equations of state. The horizontal dashed line marks 1 bar; its intersections give equilibrium densities of $0.926\pm0.007$ g/cm$^3$ for AIMD and $0.901\pm0.002$ g/cm$^3$ for DP.}
\label{figS:aimd-equation-of-state}
\end{figure}

As an independent check, liquid-water AIMD simulations were also performed in the $NVT$ ensemble at 300 K at densities of 0.80, 0.85, 0.90, 0.95, and 1.00 g/cm$^3$, following the equation-of-state approach of Ref.~\citenum{gaiduk_density_2015}. The same masses and timestep as the above $NpT$ simulations were used, with a Nos\'e--Hoover chain thermostat of chain length three and damping time 400 fs. The internal pressure $P$ was evaluated from the diagonal components $\sigma_{xx}$, $\sigma_{yy}$, and $\sigma_{zz}$ of the stress tensor along the Cartesian axes,
\begin{equation}
P=\frac{1}{3}\left(\sigma_{xx}+\sigma_{yy}+\sigma_{zz}\right),
\end{equation}
including both nuclear kinetic and electronic contributions. Figure~\ref{figS:aimd-pressure-trajectories} shows the accumulated pressure averages. A cubic fit to $P(\rho)$, shown in Figure~\ref{figS:aimd-equation-of-state}, gives an equilibrium liquid density of $0.926\pm0.007$ g/cm$^3$ at 1 bar, consistent with the direct $NpT$ result. The corresponding DP equation of state gives $0.901\pm0.002$ g/cm$^3$. DP systematically predicts a higher pressure at a given density, explaining its slightly underestimated equilibrium density.

\subsection{Test Sets from AIMD Trajectories}\label{sec:test-sets}

To test the MLIPs examined in this work, we extracted configurations from $NpT$ AIMD simulations at 300 K and 1 bar. Two test sets were constructed using plane-wave cutoffs of 400 and 800 Ry, each containing 50 liquid-water and 50 ice-Ih configurations together with their revPBE0-D3 energies and forces. The 400 Ry test set was used for BPNN1 in Ref.~\citenum{cheng_ab_2019} and NEP in Ref.~\citenum{chen_thermodynamics_2024}, matching the cutoff used to generate their training data. The 800 Ry test set was used for BPNN3 and DP trained in this work.

\section{Training-Set Density and Energy Distributions}\label{sec:coverage}

\subsection{Density Distributions}
\begin{figure}[!htb]
\centering
\includegraphics[width=0.98\textwidth]{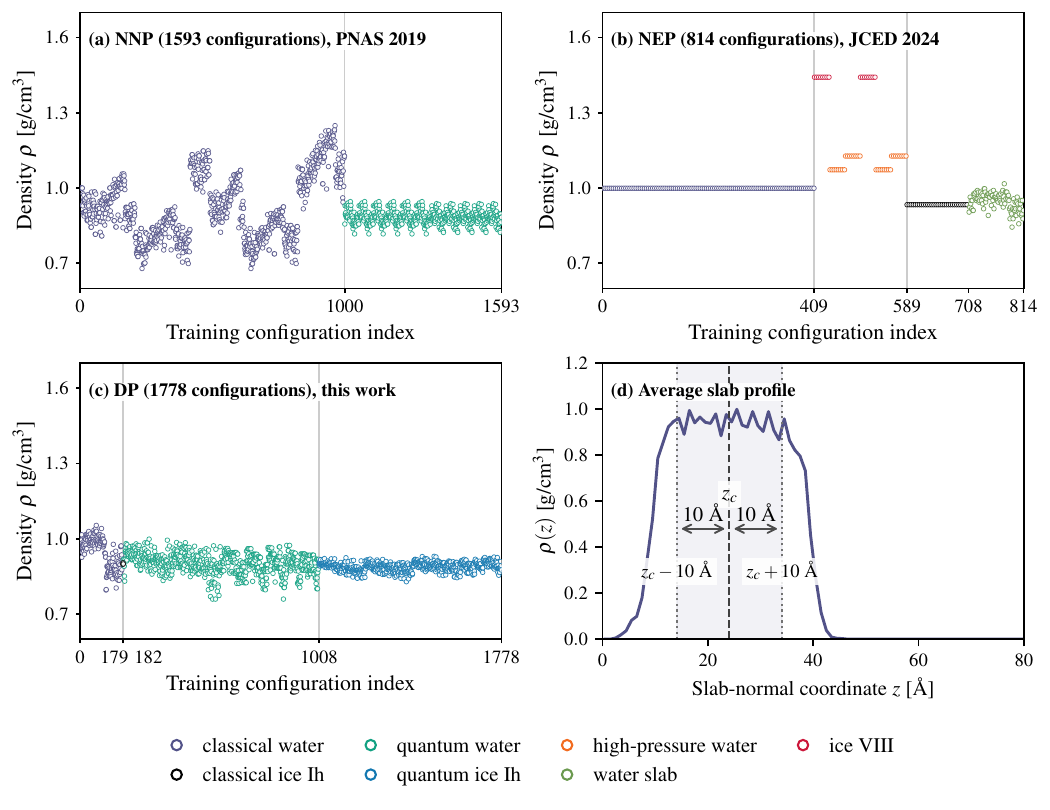}
\caption{(a--c) Configuration-by-configuration density coverage of the BPNN1 (labeled NNP), NEP, and DP training datasets for revPBE0-D3 water and ice. Colors distinguish classical water, quantum water, quantum ice Ih, high-pressure water, the water slab, and ice VIII. The NEP water-slab values are liquid-region densities obtained by averaging the bin densities within 10~\AA{} of the density-weighted slab center. For legibility, every fourth NEP non-slab configuration is plotted; all NEP slab configurations and all BPNN1 and DP configurations are shown. (d) Density profile averaged over all 106 NEP water-slab configurations using 1~\AA{} bins. The dashed line marks the density-weighted center of the average profile; dotted lines and shading indicate the central $\pm10$~\AA{} region.}
\label{figS:density-coverage}
\end{figure}

We show the per-configuration densities of the datasets discussed in the Letter in Figure~\ref{figS:density-coverage}. For NEP slabs, all cell volumes are identical. To actually report the density of their liquid regions, we exclude the vacuum region as follows. We first bin the water molecules into 1~\AA{} bins according to the oxygen positions along the slab normal and calculate a density profile, $\rho(z)$. We then find the density-weighted center of the profile, $z_c=\int z\rho(z)dz/\int \rho(z)dz$, and select a region between $z_c\pm 10$~\AA{} to calculate the liquid density by averaging $\rho(z)$ over this region. As shown in the average slab density profile in Figure~\ref{figS:density-coverage}(d), the region within $z_c\pm 10$~\AA{} covers most of the bulk liquid region. The density distributions are plotted in Figure~1 of the Letter.

\subsection{Energy Distributions}
Training-set energies in Figure~2 are DFT potential energies divided by the number of water molecules, with common bins of 0.04 eV/H$_2$O. Classical and quantum configurations retain their sampling labels; quantum configurations represent individual PIMD beads. The DP liquid distributions in Figure~2(a) were sampled in the $NpT$ ensemble at 1 bar. The classical trajectories contain 432 molecules and span 500 ps at each temperature (260, 300, 340, 400, 500, and 600 K); the quantum distribution uses a 500 ps, 32-bead PIMD trajectory of 64 molecules at 300 K. Each probability density is normalized independently using 180 bins and smoothed with a Gaussian of width two bins. The quantum distribution uses individual-bead potential energies.

\section{Instability of the SET2-Trained BPNN}\label{sec:instability}

\begin{figure}[!htb]
\centering
\includegraphics[width=0.85\textwidth]{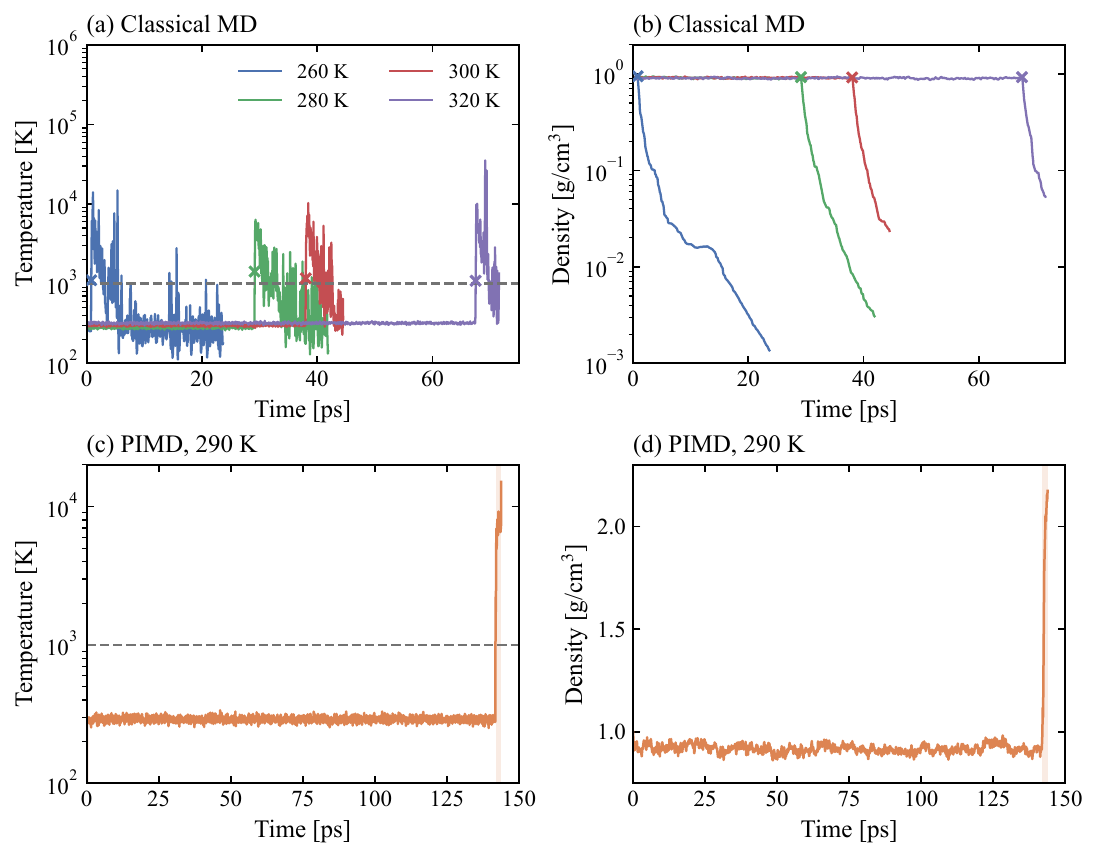}
\caption{Instability of BPNN2 trained on SET2. (a,b) Temperature and density in classical $NpT$ simulations of 432-molecule liquid water at 1 bar and the indicated target temperatures. Crosses mark the first temperature excursion above 1000 K. (c,d) Bead-normalized thermo temperature and density in a 128-molecule, 32-bead, physical-mass liquid-water PIMD simulation at 290 K and 1 bar. Orange shading begins at the first temperature excursion above 1000 K. The bead-normalized thermo temperature is shown only as a stability diagnostic and is not a thermodynamic temperature estimator.}
\label{figS:bpnn2-instability}
\end{figure}

Although BPNN2 is trained on SET2, which provides sufficient coverage for DP, it does not remain stable during extended $NpT$ simulations. Figure~\ref{figS:bpnn2-instability}(a,b) shows four classical liquid-water trajectories containing 432 H$_2$O molecules at 1 bar. At target temperatures of 260, 280, 300, and 320 K, the first temperature excursions above 1000 K occur after 0.80, 29.14, 38.02, and 67.49 ps, respectively, and are accompanied by rapid decreases in density. The 128-molecule, 32-bead PIMD trajectory at 290 K and 1 bar likewise becomes unstable after 141.89 ps, as shown in Figure~\ref{figS:bpnn2-instability}(c,d). These failures show that BPNN requires additional training configurations beyond SET2 and motivated the active-learning extension to SET3 and the training of BPNN3.

\section{Model Tests on AIMD Test Sets}\label{sec:validation}

\begin{figure}[!htbp]
\centering
\includegraphics[width=0.78\textwidth]{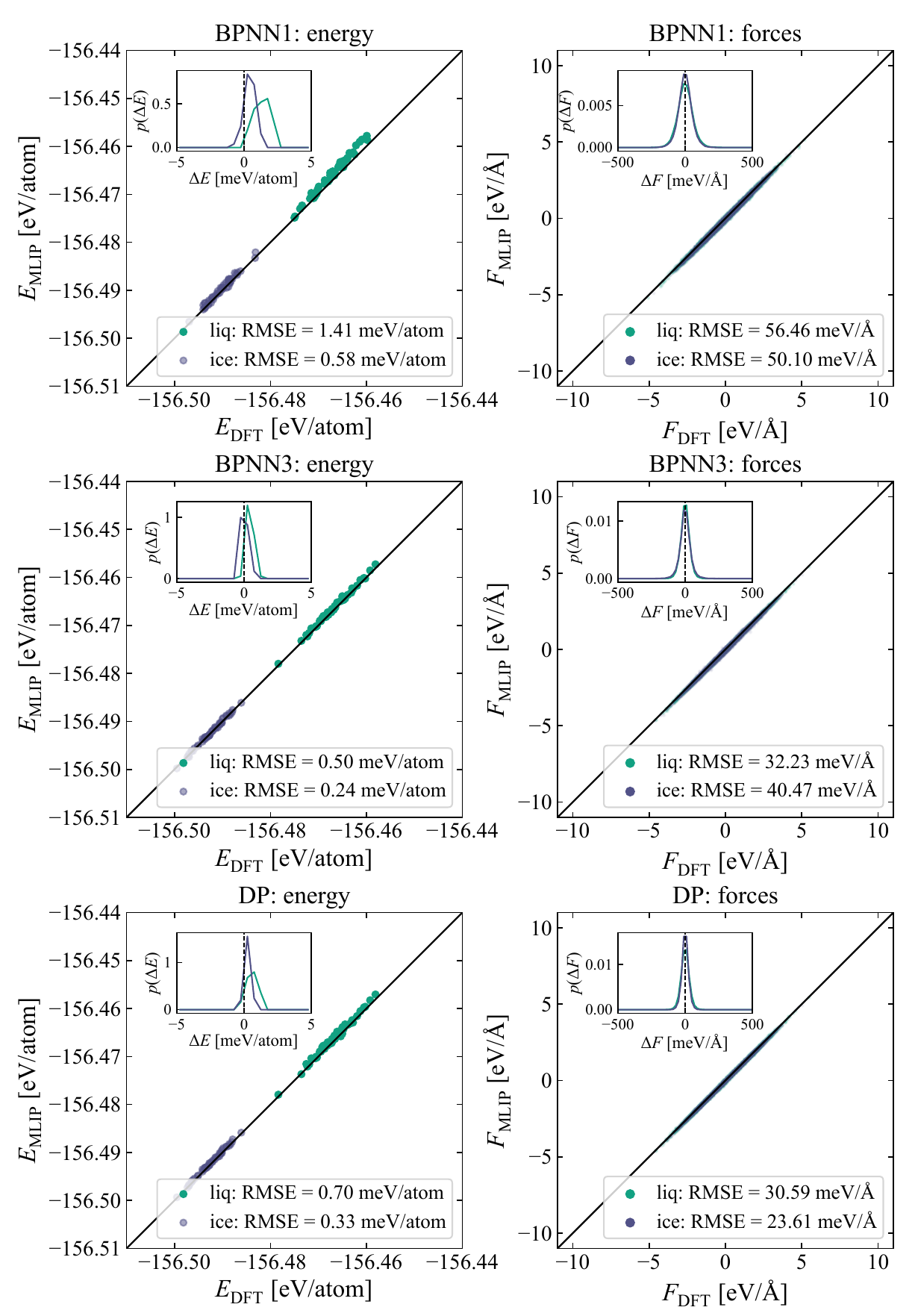}
\caption{Parity plots and error distributions for revPBE0-D3-based BPNN and DP models on AIMD configurations of liquid water and ice sampled in the $NpT$ ensemble at 300 K and 1 bar. $E$ is energy per atom and $F$ is a Cartesian force component; subscripts $\mathrm{MLIP}$ and $\mathrm{DFT}$ denote model predictions and DFT references. The errors are $\Delta E=E_{\mathrm{MLIP}}-E_{\mathrm{DFT}}$ and $\Delta F=F_{\mathrm{MLIP}}-F_{\mathrm{DFT}}$, and $p(\Delta E)$ and $p(\Delta F)$ are their probability densities. BPNN3 trained on SET3 has substantially smaller errors than the original BPNN1 trained on SET1 and performs comparably to DP on these independent configurations.}
\label{figS:test-bpnn}
\end{figure}

Figure~\ref{figS:test-bpnn} compares BPNN1, BPNN3, and DP, with energies on the left and forces on the right. BPNN3 has energy root-mean-square errors (RMSEs) of 0.50 and 0.24 meV atom$^{-1}$ and force RMSEs of 32.23 and 40.47 meV \AA$^{-1}$ for liquid water and ice, respectively. These errors are smaller than those of BPNN1 and are comparable to those of DP.

\begin{figure}[!htb]
\centering
\includegraphics[width=0.78\textwidth]{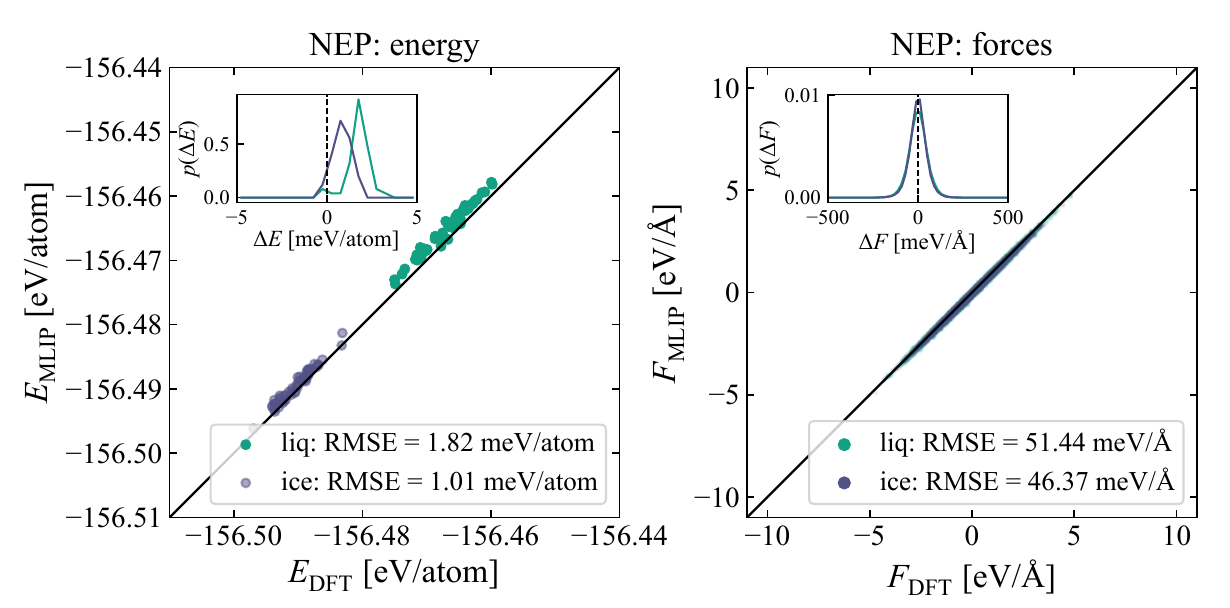}
\caption{Test errors of the revPBE0-D3-based NEP model on AIMD configurations of liquid water and ice sampled in the $NpT$ ensemble at 300 K and 1 bar.}
\label{figS:test-nep}
\end{figure}

The NEP in Ref.~\citenum{chen_thermodynamics_2024} likewise shows systematic energy biases, with energy RMSEs of approximately 1.8 meV atom$^{-1}$ for liquid water and 1.0 meV atom$^{-1}$ for ice, as shown in Figure~\ref{figS:test-nep}. These errors, together with the fixed-density sampling shown in Figures~1 and~\ref{figS:density-coverage}, suggest that additional training configurations are needed to represent ambient-pressure liquid water and ice accurately.

\section{Model Training and Active Learning}\label{sec:training}

\subsection{DP and BPNN Models}
DP models were trained using the active learning procedure described in the Letter. At each iteration, four models were trained with embedding and fitting networks containing (25, 50, 100) and (240, 240, 240) neurons, respectively. The descriptor used a 6~\AA{} cutoff, with smoothing starting at 0.5~\AA{} and 12 axis neurons. Each model was trained for 400,000 steps, with the learning rate decaying exponentially from $10^{-3}$ to $10^{-8}$ at intervals of 5000 steps. The energy and force loss prefactors varied from 0.02 to 2 and from 1000 to 1, respectively. Virials were not included in the loss function.

BPNN models were fitted to energies and forces using \texttt{n2p2}. BPNN2 and BPNN3 used atom-centered radial and angular symmetry functions with a 12-bohr cutoff, two hidden layers of 25 neurons each, hyperbolic-tangent hidden activations, and a linear output. The symmetry functions were centered and scaled using the dataset statistics. Training used a Kalman-filter optimizer for 200 epochs, with 20\% of configurations reserved for internal testing, a force weight of 1, and energy and force update fractions of 1 and 0.05, respectively. Four models with different random initializations formed the active-learning committee. The production BPNN3 model was taken at epoch 200 from the final SET3 iteration, using seed 123456.

\subsection{Dataset Composition}
SET2 was extended by BPNN active learning using the same candidate interval, $0.2\leq\mathrm{deviation}<0.35$ eV/\AA{}, and stopping criterion, more than 99.9\% of explored configurations below 0.2 eV/\AA{}, as for DP. Table~\ref{tabS:sets} summarizes the configuration counts used in Figure~2.
\begin{table}[htbp]
\centering
\caption{Training configurations by phase and nuclear treatment.}\label{tabS:sets}
\begin{tabular}{cccc}
\toprule
Configurations & SET1 & SET2 & SET3\\
\midrule
Classical liquid & 1000 & 179 & 671\\
Quantum liquid & 593 & 826 & 1023\\
Classical ice & 0 & 3 & 3\\
Quantum ice & 0 & 770 & 859\\
Total & 1593 & 1778 & 2556\\
\bottomrule
\end{tabular}
\end{table}

\section{NEP Trained on SET3}\label{sec:nep-set3}

\subsection{Model Training and Test}
A neuroevolution potential (NEP) was trained from scratch on all 2556 configurations in SET3 using the NEP3 implementation in GPUMD.\cite{fan_neuroevolution_2021,fan_gpumd_2022} Training used 400,000 generations, a batch size of 25 configurations, and a population of 52. The radial and angular cutoffs were 8 and 6~\AA{}, respectively. The input parameters were \texttt{n\_max 8 8}, \texttt{basis\_size 8 8}, and \texttt{l\_max 4 2 1}, with 50 neurons in the hidden layer. The energy and force loss weights were both 1, the virial weight was zero, and the $L_1$ and $L_2$ regularization coefficients were both 0.05. To improve numerical conditioning, a fixed energy offset $e_0=-156.392679473339$ eV/atom was removed before training, so that $E_{\mathrm{train}}=E_{\mathrm{DFT}}-N e_0$ for a configuration containing $N$ atoms. This fixed offset is restored in the energy parity plot.

\begin{figure}[!htb]
\centering
\includegraphics[width=0.78\textwidth]{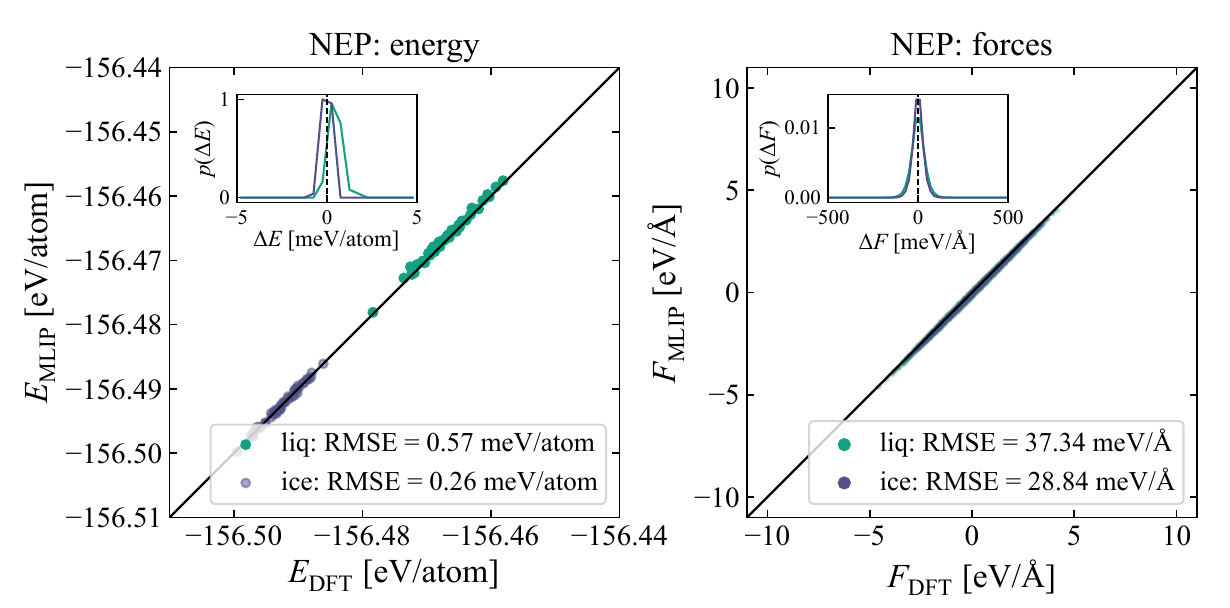}
\caption{Independent test-set energy and force parity plots for the NEP trained on SET3. Liquid-water and ice-Ih configurations were sampled from $NpT$ AIMD at 300 K and 1 bar and labeled at an 800 Ry cutoff. $E$ is the energy per atom and $F$ is a Cartesian force component. Insets show the distributions of prediction errors, $\Delta E=E_{\mathrm{MLIP}}-E_{\mathrm{DFT}}$ and $\Delta F=F_{\mathrm{MLIP}}-F_{\mathrm{DFT}}$.}
\label{figS:nep-set3-parity}
\end{figure}

Evaluation of the final potential on the full training set gives energy and force RMSEs of 0.455 meV/atom and 51.52 meV/\AA{}, respectively. Independent tests use the same 800 Ry AIMD test set as BPNN3 and DP, described in Section~\ref{sec:test-sets}. As shown in Figure~\ref{figS:nep-set3-parity}, this NEP shows a lower test error than the NEP in Ref.~\citenum{chen_thermodynamics_2024} shown in Figure~\ref{figS:test-nep} and a symmetric energy error distribution.

\subsection{Water and Ice Densities}
Classical $NpT$ simulations with this NEP were performed in GPUMD at 300 K and 1 bar using 64 H$_2$O molecules and a 0.5 fs timestep. The MTTK thermostat and barostat periods were 200 and 1000 timesteps (0.1 and 0.5 ps), respectively. Pressure coupling was isotropic for liquid water and anisotropic for ice. The liquid-water and ice trajectories lasted 2 ns and 500 ps, respectively. The first 200 ps of water and 50 ps of ice were discarded from the density estimates. Standard errors were calculated from nonoverlapping 50 ps block means.

\begin{figure}[!htb]
\centering
\includegraphics[width=0.70\textwidth]{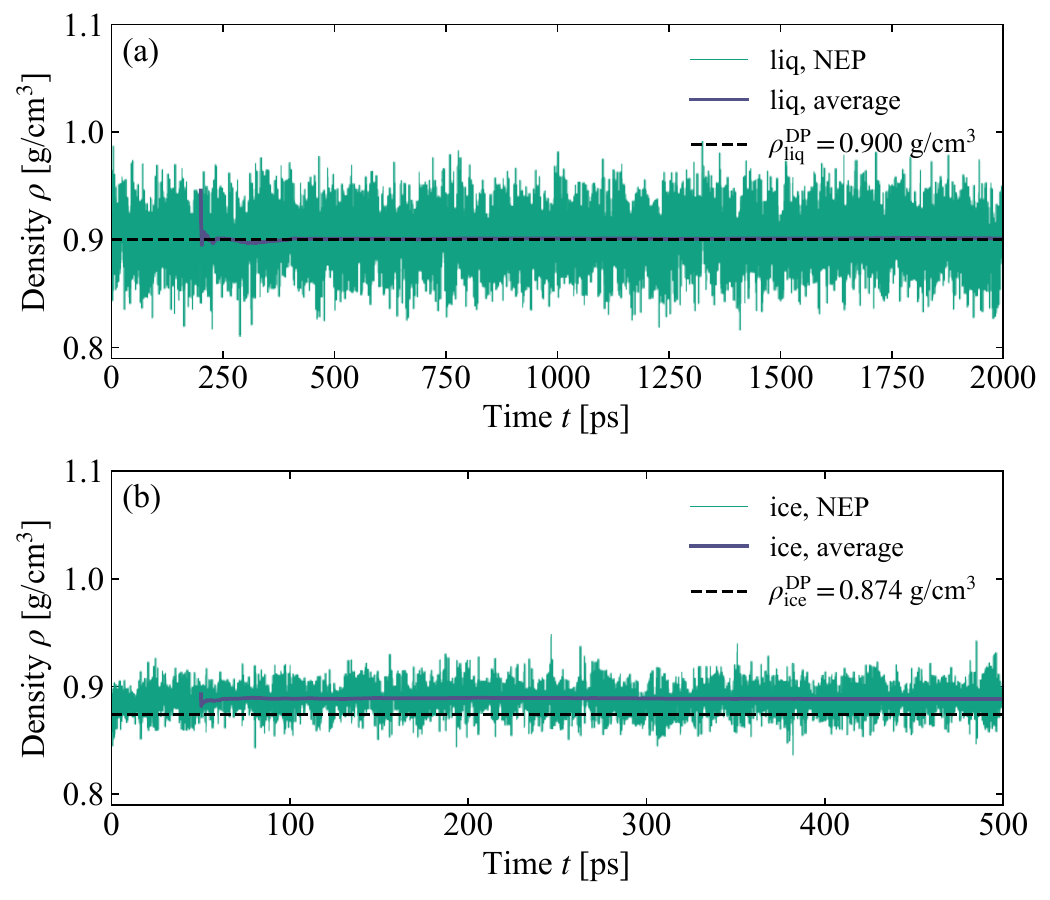}
\caption{Density trajectories of (a) liquid water and (b) ice Ih simulated with the NEP trained on SET3 at 300 K and 1 bar. Green curves show instantaneous densities, purple curves show accumulated averages after discarding the first 200 ps for water and 50 ps for ice, and black dashed lines show the corresponding 64-molecule DP mean densities. The full trajectories are shown; the panels have different time ranges.}
\label{figS:nep-set3-density}
\end{figure}

The resulting mean densities are $0.9014\pm0.0006$ g/cm$^3$ for liquid water and $0.8886\pm0.0004$ g/cm$^3$ for ice. Figure~\ref{figS:nep-set3-density} shows the trajectories and accumulated averages. The water density agrees with the corresponding DP estimate of $0.900\pm0.003$ g/cm$^3$, and the ice density is only slightly higher than the DP value of $0.874\pm0.002$ g/cm$^3$.

\section{Production MD and PIMD}\label{sec:production}

\subsection{Densities and Density Maxima}
DP simulations used LAMMPS with a 0.5 fs timestep at 1 bar. Classical MD used a Nos\'e--Hoover chain thermostat and a Martyna--Tobias--Klein barostat, with damping times of 0.1 and 0.5 ps, respectively. PIMD used the \texttt{fix pimd/langevin} module of LAMMPS, a PILE\_L thermostat, and a Bussi--Zykova--Parrinello barostat with the same damping times. Density isobars used 432 molecules and 32 beads for quantum simulations, with trajectories of 2 ns for liquid water and 0.5 ns for ice. The density maximum was located by finding the maximum of a cubic fit to the liquid density isobar, fitted after linear interpolation onto 21 evenly spaced temperatures. The 64-molecule DP comparisons to AIMD in Section~S1.2 are separate calculations.

\begin{figure}[htbp]
\centering
\includegraphics[width=0.75\textwidth]{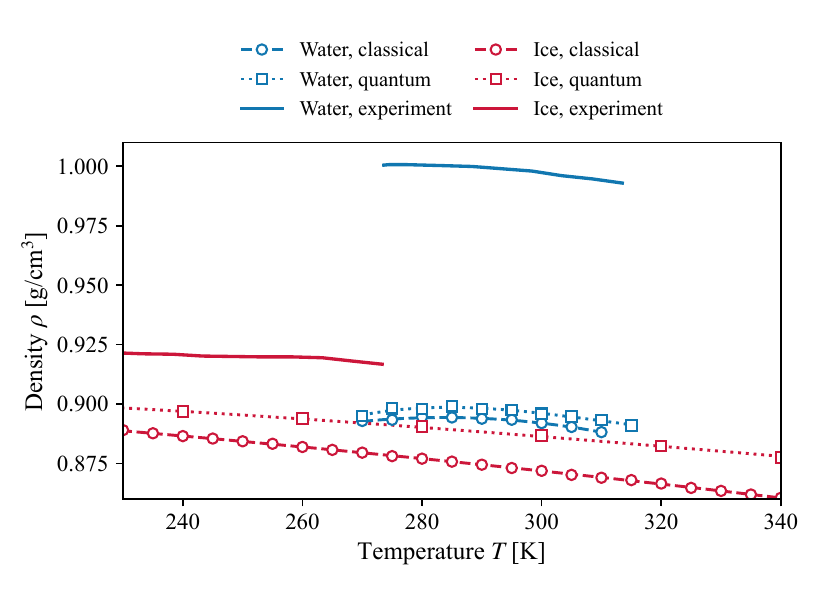}
\caption{Density isobars of liquid water and ice Ih at 1 bar. Classical and quantum DP@revPBE0-D3 results are shown by dashed and dotted curves, respectively; solid curves show experimental densities extracted from Figure~1 of Ref.~\citenum{cheng_ab_2019}, which cites Ref.~\citenum{hare_density_1987} for the supercooled liquid-water data. Symbols mark simulation results, and the DP curves are cubic fits.}
\label{figS:dp-density-isobars}
\end{figure}

Figure~\ref{figS:dp-density-isobars} shows the classical and quantum density isobars of DP@revPBE0-D3 for liquid water and ice Ih, together with experimental densities.

\subsection{Potential Energy Differences}
Figure~3 reports the mean ice--liquid potential energy differences calculated using $NpT$ trajectories at 260, 280, 300, and 320 K and 1 bar. BPNN1 and BPNN3 cells contain 64 molecules; the DP references contain 432. Each trajectory spans 500 ps, of which the first 50 ps are discarded. Mean potential energies are normalized per water molecule. Standard errors are estimated from nonoverlapping 10 ps blocks, and the uncertainty in the ice--liquid difference is obtained by adding the two phase errors in quadrature. BPNN simulations use a 0.5 fs timestep, thermostat/barostat damping times of 0.1/0.5 ps, and isotropic/anisotropic pressure coupling for liquid water/ice.

\section{Classical Free Energy Calculations}\label{sec:classical-ti}
The classical melting temperature of DP@revPBE0-D3 was obtained with the \texttt{dpti} software package; the reference systems, integration paths, and numerical procedures are detailed in Sections~2 and 3.2 of Ref.~\citenum{yuan_dpti_2026}. We used 432-molecule cells at 1 bar, with reference temperatures of 150 K for ice and 300 K for liquid water. Each reference calculation comprised 1 ns of $NpT$ equilibration, 100 ps of $NVT$ equilibration, and 500 ps per Hamiltonian-TI (HTI) window, with Simpson quadrature. Temperature integration used 500 ps per point on a 5 K grid spanning 150--350 K for ice and 250--350 K for water. Independent HTI calculations at 300 K for ice and 350 K for water verify path consistency, as shown in Figure~\ref{figS:dp-mu}(a); the crossing of the phase chemical potentials gives $\tm^{\mathrm{cl}}$. The timestep and thermostat/barostat settings follow Section~\ref{sec:production}, except that ice Hamiltonian TI uses a Langevin thermostat with a 0.1 ps damping time.

\section{Mass Thermodynamic Integration}\label{sec:mti}

\subsection{Estimator and PIMD Sampling}
PIMD simulations in MTI used 128-molecule cells at 1 bar and a 0.5 fs timestep. Physical masses of 16 and 1 atomic mass units for O and H are divided by $y^2$. The nuclear kinetic energy per molecule is evaluated with the centroid-virial estimator,
\begin{equation}
K_{\mathrm{CV}}=\frac{3N_{\mathrm{at}}k_{\mathrm B}T}{2N_{\mathrm{mol}}}
-\frac{1}{2nN_{\mathrm{mol}}}\sum_{b=1}^{n}\sum_{i=1}^{N_{\mathrm{at}}}
(\bm{r}_i^{(b)}-\bm{r}_i^{(c)})\cdot\bm{F}_i^{(b)},
\end{equation}
where $n$ is the number of beads, $N_{\mathrm{at}}$ and $N_{\mathrm{mol}}$ are the atom and molecule counts, and $\bm{r}_i^{(c)}$ is the bead centroid. Both DP and BPNN PIMD use the normal-mode OBABO integrator, a PILE\_L thermostat, and a BZP barostat, with thermostat and barostat damping times of 0.1 and 0.5 ps, respectively. Pressure coupling is isotropic for liquid water and anisotropic for ice.

Table~\ref{tabS:mti} gives the bead schedules and total trajectory lengths for Figure~4 at 290 K. The first 50 ps are discarded, and each liquid/ice pair is truncated to their common available endpoint. Standard errors are obtained from complete, nonoverlapping 20 ps blocks.
\begin{table}[htbp]
\centering
\caption{MTI bead counts and total trajectory lengths per phase before discarding equilibration for the comparison at 290 K and 1 bar.}\label{tabS:mti}
\begin{tabular}{cccccc}
\toprule
$y$ & BPNN1/3 beads & BPNN1 [ps] & BPNN3 [ps] & DP beads & DP [ps]\\
\midrule
0.1 & 8 & 350.00 & 250.00 & 8 & 500.00\\
0.2 & 16 & 350.00 & 250.00 & 16 & 500.00\\
0.3 & 32 & 346.45 & 250.00 & 32 & 500.00\\
0.4 & 32 & 350.00 & 188.25 & 64 & 500.00\\
0.6 & 32 & 350.00 & 250.00 & 64 & 500.00\\
1.0 & 32 & 229.75 & 250.00 & 64 & 500.00\\
\bottomrule
\end{tabular}
\end{table}

The comparison of BPNN1, BPNN3, and DP in Figure~4 uses 290 K and 1 bar throughout. To determine $\tm^{\mathrm{qu}}$, DP MTI was additionally evaluated at 280, 300, 310, 320, and 330 K. At every temperature, DP uses the bead schedule and total trajectory lengths in Table~\ref{tabS:mti}, a 0.5 fs timestep, and statistical errors estimated from 10 ps blocks. The reported DP integrals are calculated from the original phase-resolved kinetic-energy summaries.

\subsection{Integration and Uncertainties}
The ice--liquid integrand in Eq.~(1) of the main text is integrated with the trapezoidal rule on $y=(0,0.1,0.2,0.3,0.4,0.6,1)$, fixing the zero endpoint to zero. For independent liquid and ice estimates, its standard error is $s_g(y)=2\sqrt{s_{K,\mathrm{ice}}^2+s_{K,\mathrm{liq}}^2}/y$. The trapezoidal integral is $I=\sum_i w_i g(y_i)$, where $w_i$ are the integration weights. For the stated $y$ grid, these weights are $(0.05,0.10,0.10,0.10,0.15,0.30,0.20)$, respectively. The integral error is $s_I=[\sum_i(w_i s_{g,i})^2]^{1/2}$, assuming independent simulations at different $y$. The integrals are $0.71\pm0.08$, $-1.19\pm0.08$, and $-1.76\pm0.02$ meV/H$_2$O for BPNN1, BPNN3, and DP, respectively.

\begin{figure}[!htb]
\centering
\includegraphics[width=\textwidth]{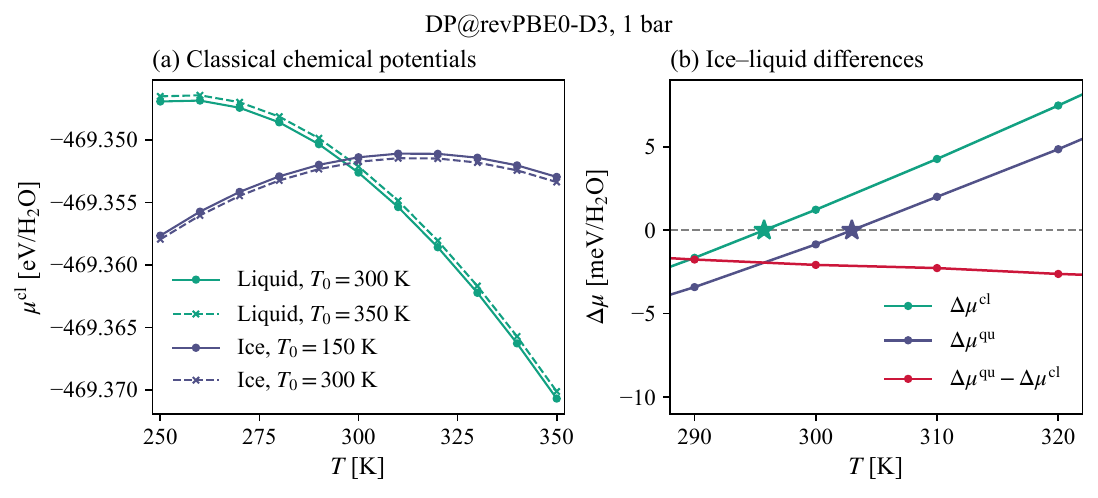}
\caption{DP@revPBE0-D3 chemical potentials at 1 bar. (a) Classical chemical potentials per molecule obtained from TI with the reference temperatures $T_0$. (b) Classical and quantum ice--liquid differences, $\Delta\mu=\mu_{\mathrm{ice}}-\mu_{\mathrm{liq}}$, and their NQE correction. The classical baseline in panel (b) uses HTI reference temperatures of 150 K for ice and 300 K for liquid water. Stars mark the zero crossings of the classical and quantum curves.}
\label{figS:dp-mu}
\end{figure}

Figure~\ref{figS:dp-mu}(b) shows the classical and quantum ice--liquid chemical-potential differences. Their zero crossings give $\tm^{\mathrm{cl}}\simeq296$ K and $\tm^{\mathrm{qu}}\simeq303$ K.

\bibliography{clean}